\documentclass[11pt]{article}
\usepackage{a4wide}
\usepackage{epsfig}
\usepackage{amssymb}
\usepackage{amsmath}
\usepackage{amsthm}
\usepackage{multirow}

\newcommand{\beq}{\begin{equation*}}
\newcommand{\eeq}{\end{equation*}}
\newcommand{\beqa}{\begin{eqnarray*}}
\newcommand{\eeqa}{\end{eqnarray*}}

\newcommand{\argmin}{\mathop{\rm arg\, min}}

\newcommand{\R}{\mathbb{R}}
\newcommand{\E}{\mathrm{E}}

\newcommand{\T}{\mathcal{T}}
\newcommand{\F}{\mathcal{F}}

\newcommand{\bh}{\hat{\beta}}
\newcommand{\be}{\beta}
\newcommand{\bm}{\beta_{{\rm min}}}
\newcommand{\bt}{\beta_{\T}}
\newcommand{\btf}{\be_{\T-\F}}
\newcommand{\Xt}{X_{\T}}
\newcommand{\Xbtf}{X_{\T-\F}\be_{\T-\F}}
\newcommand{\e}{\epsilon}

\newcommand{\Ss}{\mathcal{S}}

\newcommand{\eq}{\begin{equation*}}
\newcommand{\en}{\end{equation*}}
\newcommand{\eqn}{\begin{equation}}
\newcommand{\enn}{\end{equation}}
\newcommand{\eqa}{\begin{eqnarray}}
\newcommand{\ena}{\end{eqnarray}}

\long\def\comment#1{}

\newtheorem{thm}{Theorem}

\newtheorem{lem}[thm]{Lemma}
\newtheorem{cor}[thm]{Corollary}

\title{Sharp Sufficient Conditions on Exact Sparsity Pattern Recovery}

\author{Kamiar Rahnama Rad
\thanks{Kamiar Rahnama Rad is with the Department
of Statistics, Columbia University, New York,
NY, 10027 USA. e-mail: kamiar@stat.columbia.edu.} 
\thanks{Part of this work was
presented at the 43rd Annual Conference on Information Sciences and Systems in March 2009.}
}

\begin{document}

\maketitle
\begin{abstract}
Consider the $n$-dimensional vector 
$y=X\be+\e$, where $\be \in \R^p$ has only $k$ nonzero entries 
and $\e \in \R^n$ is a Gaussian noise.
This can be viewed as a linear system with sparsity 
constraints, corrupted by noise. We find a non-asymptotic 
upper bound on the probability that the optimal decoder 
for $\beta$ declares a wrong sparsity pattern,
given any generic perturbation matrix $X$. 
In the case when $X$ is randomly drawn from 
a Gaussian ensemble, we obtain asymptotically sharp 
sufficient conditions for exact recovery,
which agree with the known necessary conditions 
previously established.
\end{abstract}

\section{Introduction}
A wide array of problems in science and 
technology reduce to finding solutions
to underdetermined systems of equations, 
particularly to systems of linear
equations with fewer equations than unknowns;
examples include array signal processing~\cite{ZP01}, neural~\cite{VG00} 
and genomic data analysis~\cite{B05}, to name a few. 
In many of these applications, it is natural to
seek for {\em sparse} solutions of
such systems, i.e., solutions with few nonzero elements.
A common setting is when we believe or we know {\em a priori} 
that only a {\em small subset} of the candidate  sources, 
neurons, or genes influence the observations,
but their location is unknown.

More concretely, the problem we consider is that
of estimating the support of $\be \in \R^p$,
given the {\em a priori} knowledge that only $k$ of its entries 
are nonzero, and based on the following observational model,
\begin{equation}
y=X\be+\e,
\label{eq:model}
\end{equation}
where $X \in \R^{n \times p}$ is a collection of perturbation vectors, 
$y \in \R^n$ is the output measurement and $\e \in \R^n$ is the 
additive measurement noise, assumed to be zero mean and 
with known covariance equal to $I_{n \times n}$; 
this entails no loss of generality, by standard rescaling 
of $\beta$. Each row of $X$ and the corresponding entry of $y$ are viewed as an input perturbation and output measurement, respectively.  
For that reason, $n$ designates the size of {\em measurements}, $p$ size of {\em features} and $k$ size of {\em relevant features}. As mentioned earlier, the main problem is to optimally estimate the set of nonzero entries of $\beta$, i.e. the {\em sparsity pattern}, based on the $n$-dimensional
observation vector $y$ and the 
$(m\times n)$ perturbation matrix $X$,
and to study conditions on the key parameters that 
guarantee (asymptotically) that 
the sparsity pattern is recovered reliably. 
The geometric structure of the problem  is represented by $p$ and $k$,
whereas the size of the measurements and signal-to-noise ratio are given by $n$ 
and $\| \be\|^2_2$, respectively. Therefore,
$(n,p,k,\|\be\|_2^2)$ may be viewed as the key parameters that
asymptotically determine whether reliable sparsity pattern recovery 
is possible or not. The aforementioned question can be posed 
in terms of $(n,p,k,\bm^2)$, where $\bm=\min_{i}|\be_i|$,
upon noting that $\|\be\|^2_2 \geq k \bm^2$.

A large body of recent work, including 
\cite{WWR08,Wain07,AT08,FRG08, Karbasi09},
analyzed reliable sparsity pattern recovery exploiting optimal and sub-optimal decoders for large random Gaussian perturbation matrices. The average error probability, necessary and sufficient conditions for sparsity pattern recovery for Gaussian perturbation matrices were analyzed in ~\cite{Wain07}. As a generalization of the previous work, necessary conditions for general random and sparse perturbation matrices  were presented in ~\cite{WWR08}. Various performance metrics regarding the sparsity pattern estimate were examined in~\cite{AT08}. We will discuss the relationship to this
work below in more depth, after describing our analysis and results in more detail. 

The output of the optimal (sparsity) decoder 
is defined as the support set of the sparse 
solution $\hat{\beta}$ with support size $k$ that 
minimizes the residual sum of squares, where,
\begin{equation}
\bh = \argmin_{ |{\rm support}(\theta)|=k} \|y-X\theta \|_2^2 
\label{eq:decoder},
\end{equation}
is the optimal estimate of $\beta$ given the {\em a priori}
information of sparseness. The support set of $\bh$ is optimal in 
the sense of minimizing the probability of 
identifying a wrong sparsity pattern. 

Below, first, we present an upper bound on the probability of declaring a wrong sparsity pattern based on the optimum decoder,
as a function of the perturbation matrix $X$. 
Second, we exploit this upper bound to find asymptotic sufficient 
conditions on $(n,p,k,\bm^2)$ for reliable sparsity recovery,
in the case when the entries of the perturbation matrix are 
independent and identically distributed (i.i.d.)
normal random variables. Finally, we 
show that our results strengthen earlier sufficient conditions 
\cite{Wain07,Karbasi09,AT08,FRG08}, and we establish the sharpness 
of these sufficient conditions in both the linear, 
i.e., $k=\Theta(p)$, and the sub-linear, 
i.e., $k=o(p)$, regimes,
for various scalings of $\bm^2$.  

\medskip

\textbf{Notation. } The following conventions will remain in 
effect throughout this paper. Calligraphic letters are used to indicate sparsity patterns defined as a set of integers between 1 and $p$, with cardinality 
$k$.   We say $\be \in \R^p$ has sparsity pattern $\T$  if only entries with indices $i \in \T$ are nonzero. $\T-\F$ stands for the set of entries that are in $\T$ but not in $\F$ and $|\T|$ for the cardinality of $\T$. We generally denote by $X_\T \in \R^{n \times |\T|}$, the matrix obtained from $X$ by extracting $|\T|$ columns with indices obeying $i \in \T$. 
Let $\Ss(\be)$ stand for the sparsity pattern or 
support set of $\be$. All norms are $\ell_2$,
$\|\cdot\|=\|\cdot\|_2$.

\subsection{Results}
For the observational model in equation (\ref{eq:model}),
assume that the true sparsity model is $\T$,  so that,
\begin{equation}
y=\Xt\bt+\e. \label{eq:model-1}
\end{equation}
We first state a result on the probability
of the event $\Ss(\bh)=\F$,
for any $\F \neq \T$ and any perturbation matrix $X$. 
\begin{thm}
For the observational model of equation (\ref{eq:model-1}) and estimate $\bh$ in equation (\ref{eq:decoder}),
the conditional  probability $\Pr[\Ss(\bh)=\F| X,\beta,\T]$ that the decoder declares $\F$ when $\T$ is the true sparsity pattern, is bounded above 
by $e^{-c\|(I-\Pi_\F)X_{\T-\F}\beta_{\T-\F} \| + \frac{d}{2}}$, where $c=\frac{3-2\sqrt{2}}{2}$, $d=|\T-\F|$ and $\Pi_\F=X_\F(X_\F^TX_\F)^{-1}X_\F^T$. \label{thm1}
\end{thm}
The proof of Theorem \ref{thm1}, given in Section \ref{thm1-proof}, employs the Chernoff technique and the properties of the eigenvalues of the difference of projection matrices, to bound the probability of declaring a wrong sparsity pattern $\F$  instead of the true one $\T$ as function of the perturbation matrix $X$ and the true parameter $\beta$.
The error rate decreases exponentially in the norm of the projection of $X_{\T-\F}\beta_{\T-\F}$ on the orthogonal subspace spanned by the columns of $X_\F$. This is in agreement with the intuition that, the closer different subspaces corresponding to different sets of columns of $X$ are, the harder it is to differentiate them, and hence the higher the error probability will be.


The theorem below gives a non-asymptotic bound on the probability
of the event $\Ss(\bh)\neq\T$, when the entries of the perturbation 
matrix $X$ are drawn i.i.d.\ from a normal distribution.

\begin{thm} \label{thm2} For the observational model 
of equation (\ref{eq:model-1}) and the estimate $\bh$ in 
equation (\ref{eq:decoder}),
if the entries of $X$ are i.i.d.\ $\mathcal{N}(0,1)$, $p>2k$,
\begin{equation}
(n-k)\bm^2 > 4 \frac{(1+k\bm^2)^2}{k\bm^2} \label{eq:conv-cond-1},
\end{equation} 
and
\begin{equation*}
n - k>  C \max \left \{\frac{\log k (p-k)}{\log(1+\bm^2)}, \frac{k\log(\frac{p-k}{k}) + \log k}{\log(1+k\bm^2)} \right \},\\
\end{equation*}
 then 
 \begin{equation*}
 \Pr[\Ss(\bh) \neq \T] \leq k e^{5/2} \max \left \{ (p-k)^{-B},  \left[ \frac{e(p-k)}{k} \right]^{-kB} \right \},
\end{equation*}
for   $B=\frac{C-5}{2}$.
\end{thm}

The proof of Theorem \ref{thm2}, given in Section \ref{thm2-proof}, uses union bound together with counting arguments similar in spirit to those \cite{Wain07},
to bound the probability
of error of the optimal decoder.  
 
If we let $n(p)$, $k(p)$ and $\bm(p)$ scale as a function of 
$p$, then the upper bound of $\Pr[\Ss(\bh) \neq \T]$ scales 
like $k (p-k)^{-B}$. For $B >2$ or, equivalently, $C >9$ 
the probability of error as $p \rightarrow \infty$ is 
bounded above by $p^{-D}$ for some $D>1$. 
Therefore, the following sum,
\begin{equation}
\sum_{p=1}^{\infty} \Pr[\Ss(\bh_{p \times 1}) \neq \T_p],
\end{equation}
is finite, and as a consequence of Borel-Cantelli lemma, 
for large enough $p$,  the decoder declares the true sparsity 
pattern almost surely. In other words, the estimate $\bh$  based on (\ref{eq:decoder}) achieves the same loss as an oracle which is supplied 
with perfect information about which coefficients of $\beta$ are nonzero. 
The following corollary summarizes the aforementioned statements.
\begin{cor} \label{cor}
For the observational model  of equation (\ref{eq:model-1})  
and the estimate $\bh$ in equation (\ref{eq:decoder}),
let $n$, $k$ and $\bm^2$ scale as a function of $p$,
such that $ (n-k)\bm^2 > 4 \frac{(1+k\bm^2)^2}{k\bm^2}$. 
Then there exists a constant $C^{\star}$ such that, if 
\begin{equation*}
n > C^{\star} \max \left \{\frac{\log(p-k)}{\log{(1+\bm^2)}}, \frac{k \log(\frac{p}{k})}{\log(1+k \bm^2)} ,k \right \} ,
\end{equation*}
 then  a.s.  for large enough $p$, $\bh$ achieves the same performance loss 
as an oracle which is supplied with perfect information about which coefficients of $\beta$ are nonzero and $\Ss(\bh) = \T$.
\end{cor}

The sufficient conditions in Corollary \ref{cor} can be compared 
against similar conditions for exact sparsity pattern recovery in 
\cite{Wain07, FRG08,AT08,Karbasi09}; for example, in the sub-linear regime 
$k=o(p)$, 
when $\bm^2=\Theta(1)$, \cite{Wain07,Karbasi09} proved 
that $n=\Theta(k \log(\frac{p}{k}))$ is sufficient,
and \cite{AT08,FRG08} proved that $n=\Theta(k \log(p-k))$ 
is sufficient. In that vain, according to Corollary \ref{cor},
\begin{eqnarray*}
n&=&\max  \left \{ \Theta \left (\frac{k \log (\frac{p}{k})}{\log k} \right),\Theta(k) \right \},
\end{eqnarray*}
suffices to ensure exact sparsity pattern recovery
and, therefore, it  strengthens these earlier results.


\begin{table}[ht!]
\begin{tabular}[c]{| l || c | r |} 
\hline
  Scaling             						& Sufficient condition        							  									&          Necessary condition    \\ 
                          						& Corollary 3                    								  									&          Theorem 4 \cite{WWR08}\\ \hline \hline
  $k=\Theta(p)$ 						& 														  									&	      \\
  $\bm^2=\Theta(\frac{1}{k})$  		& $n=\Theta(p \log p)$          							  									& 		 $n=\Theta(p \log p)$  \\ \hline
  $k=\Theta(p)$ 						& 														  									&	       \\
  $\bm^2=\Theta(\frac{\log k}{k})$       & $n=\Theta(p)$		          							  									& 		 $n=\Theta(p)$\\ \hline
  $k=\Theta(p)$ 						& 														  									&	      \\
  $\bm^2=\Theta(1)$  					& $n=\Theta(p)$               								  									& 		 $n=\Theta(p)$   \\ \hline \hline
  $k=o(p)$ 								& 														  									&	       \\
  $\bm^2=\Theta(\frac{1}{k})$  		& $n=\Theta(p \log (p-k))$          				      	  									& 		 $n=\Theta(p \log (p-k))$  \\ \hline
  $k=o(p)$ 								& 													        									&	       \\
  $\bm^2=\Theta(\frac{\log k}{k})$  	& $n=\Theta \left(\frac{k \log (\frac{p}{k})}{\log \log k} \right)$      							& 		 $n=\Theta(\frac{k \log (\frac{p}{k})}{\log \log k})$ \\ \hline
  $k=o(p)$ 								& 														  									&	       \\
  $\bm^2=\Theta(1)$			  		& $n=\max\left\{\Theta\left(\frac{k\log(\frac{p}{k})}{\log k}\right),\Theta(k) \right \}$        	&$n=\max\left\{\Theta\left(\frac{k\log(\frac{p}{k})}{\log k}\right),\Theta(k) \right \}$    \\ \hline 
\end{tabular}
\caption{Tight necessary and sufficient conditions on the number of measurements $n$
required for reliable support recovery in different regimes of interest. }
\label{table}
\end{table}

What remains is to see whether the sufficient conditions in Corollary \ref{cor} match the necessary conditions proved in \cite{WWR08} : 
\begin{thm} \label{thm-wang}
\cite{WWR08}: Suppose that the entries of the perturbation 
matrix $X \in \R^{n \times p}$ are drawn i.i.d.\
from any distribution with zero-mean and variance one. 
Then a necessary condition for
asymptotically reliable recovery is that:
\begin{equation*}
n > \max \{ f_1(k,p,\bm^2),f_2(k,p,\bm^2),k-1\},
\end{equation*}
where
\begin{eqnarray*}
f_1(k,p,\bm^2) &=&  \frac{\log {p \choose k} -1}{\frac{1}{2} \log (1+k\bm^2(1-\frac{k}{p}))}\\
f_2(k,p,\bm^2) &=&  \frac{\log (p-k +1) -1}{\frac{1}{2} \log (1+\bm^2(1-\frac{1}{p-k+1}))}.
\end{eqnarray*}
\end{thm}


The necessary condition in Theorem \ref{thm-wang} asymptotically 
resembles the sufficient condition in Corollary \ref{cor}; 
recall that $\log {p \choose k} < k \log (\frac{ep}{k})$. 
The sufficient conditions of Corollary \ref{cor} can be compared 
against the necessary conditions in \cite{WWR08} for exact sparsity 
pattern recovery, as shown in Table \ref{table}. We obtain tight 
sufficient conditions which match the necessary conditions 
in the regime of linear and sub-linear signal sparsity, under 
various scalings of the minimum value $\bm$.

\section{Proof of Theorems}
\subsection{Theorem 1} \label{thm1-proof}
For a given sparsity pattern $\F$,
the minimum residual sum of squares is achieved by,
\begin{eqnarray*}
\min_{\theta_\F \in \R^k} \|y-X_\F \theta_\F\|^2 &=& \| y - \Pi_\F y\|^2,
\end{eqnarray*}
where $\Pi_\F$ denotes the orthogonal projection operator into 
the column space of $X_\F$; among all sparsity patterns with 
size $k$, the optimum decoder declares,
\begin{equation*}
\hat{\T}(y,X) = \argmin_{|\F|=k} \| y - \Pi_\F y\|^2,
\end{equation*}
as the optimum estimate of the true sparsity pattern 
in terms of minimum error probability. 
Recall the definition of $\bh$ in equation (\ref{eq:decoder}) and note that $\Ss(\bh)=\hat{\T}(y,X)$. It is clear that the decoder 
incorrectly declares $\F$ instead of the true sparsity pattern
(namely $\T$), if and only if,
\begin{equation*}
\| y - \Pi_\F y\|^2 < \| y - \Pi_\T y\|^2,
\end{equation*}
or equivalently,
\begin{equation*}
Z_\F:=y^T(\Pi_\F-\Pi_\T)y > 0.
\end{equation*}
The rest of the proof reduces
to finding an upper bound on the probability that 
$Z_\F >0$ with the aid of the Chernoff technique:
\begin{eqnarray*}
\Pr[Z_\F > 0| X,\T,\beta] 
&\leq& \inf_{|t|<1/2} \E[e^{Z_\F t} | X,\T,\beta] \\
&\leq& e^{-c\|(I-\Pi_\F)\Xbtf \|^2 + \frac{d}{2}}.
\end{eqnarray*}
The infimum is taken over $|t|<1/2$ to guarantee boundedness of the expectation. The last inequality, proven in the next lemma, concludes the proof.
\begin{lem}
For $y \sim \mathcal{N}(\Xt\bt,I)$ define $Z=y^T (\Pi_\F-\Pi_\T)y$ 
and let $|\F-\T|=d$. then:
\begin{equation*}
\inf_{|t|<1/2}\log \E[e^{Zt}|X,\T,\beta] \leq \frac{d}{2}-\frac{3-2\sqrt{2}}{2} \| (I-\Pi_\F)\Xbtf\|^2.
\end{equation*} 
\end{lem}
\textbf{Proof.} Note that for $y \sim \mathcal{N}(\mu,I)$ Gaussian integrals yield:
\begin{eqnarray*}
\E[e^{ty^T\Psi y}] &=& (2\pi)^{-\frac{n}{2}}\int e^{t(\mu+\e)^T\Psi(\mu+\e)} e^{-\frac{\| \e\|^2}{2} } d\e \\
&=& \frac{e^{t \mu^T\Psi\mu+2t^2\mu^T\Psi(I-2t\Psi)^{-1}\Psi\mu}}{\det(I-2t\Psi)^{\frac{1}{2}}} \int \frac{e^{-\frac{\|(I-2t\Psi)^{1/2}(\e-\e_0) \|^2}{2}}}{(2\pi)^{n/2}\det(I-2t\Psi^{-\frac{1}{2}})}d\e,
\end{eqnarray*}
where $\e_0=2t(I-2t\Psi)^{-1}\Psi\mu$.  Thus,
\begin{equation*}
\log \E [e^{Zt}] = 2t^2\mu^T\Psi (I-2t\Psi)^{-1}\Psi \mu + t\mu^T\Psi\mu-\frac{1}{2}\log \det(I-2t\Psi).
\end{equation*}
Substituting $\mu=\Xt\bt$ and $\Psi=\Pi_\F-\Pi_\T$ we obtain,
\begin{eqnarray}
\mu^T\Psi\mu &=& - \| (I-\Pi_\F)\Xt \bt\|^2 \nonumber \\
&=& - \| (I-\Pi_\F)\Xbtf\|^2,\label{eq:mu1}
\end{eqnarray}
and similarly, we have,
\begin{equation}\label{eq:mu2}
\mu^T \Psi^2\mu = \|(I-\Pi_\F) \Xbtf\|^2.
\end{equation}

Therefore,
\begin{eqnarray}
\log \E[ e^{Zt}] &=& 2t^2   \mu^T \Psi (I-2t\Psi)^{-1} \Psi \mu + t\mu^T\Psi\mu-\frac{1}{2}\log \det(I-2t\Psi) \nonumber \\
&\stackrel{1}{\leq}&  2t^2 \| (I-2t\Psi)^{-1/2}\|^2   \mu^T \Psi^2 \mu + t\mu^T\Psi\mu-\frac{1}{2}\log \det(I-2t\Psi) \nonumber \\
&\stackrel{2}{=}& \left \{ 2t^2 \| (I-2t\Psi)^{-1/2} \|^2    - t \right \}\|(I-\Pi_\F) \Xbtf\|^2 -\frac{1}{2}\log \det(I-2t\Psi) \nonumber \\
&\stackrel{3}{\leq}& \left [ 2t^2 \| (I-2t\Psi)^{-1/2} \|^2 - t \right ]  \|(I-\Pi_\F) \Xbtf\|^2  -\frac{d}{2} \log (1-4t^2)  \nonumber \\
&\stackrel{4}{\leq}&\left [ \frac{2t^2}{1-2t} - t \right ]  \|(I-\Pi_\F) \Xbtf\|^2   -\frac{d}{2} \log (1-4t^2). \label{eq:cdf-bound}
\end{eqnarray}
The first inequality follows by an application
of the Cauchy-Schwarz inequality and the second equality follows from equations (\ref{eq:mu1},\ref{eq:mu2}). Regarding the third and fourth inequality note that the top eigenvalue of $\Psi=\Pi_\F-\Pi_\T$ is bounded by one and therefore $I-2t\Psi$ is positive definite for $|t|<1/2$.  The difference of projection matrices $\Pi_\F-\Pi_\T$ has $d=|\T-\F|$ pairs
of nonzero positive and negative eigenvalues, bounded above by one 
and bounded below by negative one, respectively, and equal in magnitude. 
Letting the $d$ positive eigenvalues of $\Pi_\F-\Pi_\T$ be denoted by $\lambda_1,\cdots,\lambda_d$,
\begin{eqnarray*}
\log \det (I-2t \Psi) &=& \sum_{i=1}^d \{ \log(1-2t\lambda_i)+\log(1+2t\lambda_i) \} \\
&=& \sum_{i=1} ^d\log(1-4t^2\lambda_i^2)\\
&\geq& d \log(1-4t^2).
\end{eqnarray*}
Furthermore, 
\begin{eqnarray*}
\| (I-2t\Psi)^{-1/2}\|^2 &=& \max_{1\geq i \geq d} (1-2t\lambda_i)^{-1} \\
&\leq& (1-2t)^{-1},
\end{eqnarray*}
 which yields the fourth inequality.
 Finally,  since inequality (\ref{eq:cdf-bound}) is true for any $|t|<1/2$ we take the infimum of  $\frac{2t^2}{1-2t} - t$ over $|t|<1/2$ which is equal to $\sqrt{2}-3/2$ at $t=1/2(1-\sqrt{2}/2)$ and obtain the desired bound:
 \begin{eqnarray*}
\inf_{|t|<1/2} \log \E [e^{Zt}] &\leq& -\frac{3-2\sqrt{2}}{2}  \|(I-\Pi_\F) \Xbtf\|^2   -\frac{d}{2} \log (\sqrt{2}-1/2) \\
 &\leq&  -\frac{3-2\sqrt{2}}{2}  \|(I-\Pi_\F) \Xbtf\|^2   +\frac{d}{2}.
 \end{eqnarray*}

\subsection{Theorem 2} \label{thm2-proof}
First, to find conditions under which $\Pr[E_p]$  asymptotically goes to zero, 
with $E_p$ defined as the event that $\Ss(\bh)$ is not equal to $\T$, we exploit the union bound in conjunction with counting arguments and lemma \ref{lem-thm2} proved below. We have:
\begin{eqnarray}
\Pr[E_p] &=& \Pr \left [\cup_{\F\neq\T} \{ Z_\F >0\} \right ]  \nonumber \\
&\leq& \sum_{\F \neq \T} \Pr \left [  Z_\F >0  \right ] \nonumber \\
&=& \sum_{d=1}^k \sum_{|\F-\T|=d} \Pr \left [  Z_\F >0  \right ] \nonumber \\
&\stackrel{1}{=}& \sum_{d=1}^k \sum_{|\F-\T|=d}  e^{-\frac{n-k}{2} \log(1+2c\|\btf \|^2) + \frac{d}{2}} \nonumber \\
&\stackrel{2}{\leq}& \sum_{d=1}^k {k \choose d} {p-k \choose d}  e^{-\frac{n-k}{2} \log(1+2c d \bm^2) + \frac{d}{2}} \nonumber \\
&\stackrel{3}{\leq}& \sum_{d=1}^k  e^{d[\frac{5}{2} +\log(\frac{k(p-k)}{d^2})]-\frac{n-k}{2} \log(1+2c d \bm^2)} \nonumber \\
&\leq& k  e^{\max \left \{ \frac{5}{2} +\log(k(p-k))-\frac{n-k}{2} \log(1+2c  \bm^2), k[\frac{5}{2} +\log(\frac{p-k}{k})]-\frac{n-k}{2} \log(1+2c k \bm^2) \right \} }\label{eq:err-bound}
\end{eqnarray}
The first inequality is proved in Lemma \ref{lem-thm2} below,
and the second inequality follows from the observation that 
there are ${k \choose d} {p-k \choose d}$ sparsity patterns that 
differ in exactly $d$ elements with $\T$. For the third inequality recall the definition of $\bm$ and that $\log {a \choose b} < b \log(\frac{ae}{b}) $. Finally, the last inequality follows from the convexity of 
the function,
\begin{equation*}
f(d):= d[\frac{5}{2} +\log(\frac{k(p-k)}{d^2})]-\frac{n-k}{2} \log(1+2c d \bm^2),
\end{equation*}
when,
\begin{equation}
(n-k)\bm^2 >  4 \frac{(1+k\bm^2)^2}{k\bm^2}
 \label{eq:conv-cond}.
\end{equation} 
As a consequence of convexity the maximum of $f(.)$ is attained at its boundary which is $d=1$ and $d=k$.
To see that $f(d)$ is convex, taking derivatives yields,
\begin{eqnarray*}
f'(d) &=& \frac{5}{2} +\log(\frac{k(p-k)}{d^2})-\frac{c \bm^2(n-k)}{1+2c d \bm^2}\\
f''(d) &=& -\frac{2}{d} + \frac{2c^2 \bm^4(n-k)  }{(1+2c d \bm^2)^2}.
\end{eqnarray*}
and inequality (\ref{eq:conv-cond}) yields  $f''(d)>0$.  
Therefore, for $\Pr[E_p] \rightarrow 0$, it suffices that,
\begin{eqnarray}
n-k &>& C \max \left \{\frac{\log (p-k)}{\log(1+\bm^2)}, \frac{k\log(\frac{p-k}{k}) +k}{\log(1+k\bm^2)} \right \},\label{eq:cond}
\end{eqnarray}
for a large enough constant $C$.
Now, given condition (\ref{eq:cond}) above, we obtain a 
non-asymptotic upper bound on the error probability by continuing from equation (\ref{eq:err-bound}). To this end we have,
\begin{eqnarray}
\frac{5}{2} +\log(k(p-k))-\frac{n-k}{2} \log(1+2c  \bm^2) &\leq& \frac{5}{2} + \log(k(p-k))- \frac{C}{2} \log(p-k) \nonumber \\
&\leq&  \frac{5}{2}  - \frac{C-5}{2}  \log(p-k) \label{eq:exp1},
\end{eqnarray}
since $2k<p$, and similarly,
\begin{eqnarray}
 k\left [\frac{5}{2} +\log(\frac{p-k}{k}) \right]-\frac{n-k}{2} \log(1+2c k \bm^2) &\leq&  k \left [\frac{5}{2} +\log(\frac{p-k}{k}) \right ]-\frac{C}{2} \left [k \log(\frac{p-k}{k}) + k \right ] \nonumber \\
 &\leq& - \frac{C-5}{2} \left [  k \log (\frac{p-k}{k}) + k \right ] \label{eq:exp2}.
\end{eqnarray}
In the end, if inequality (\ref{eq:cond}) is satisfied,  inequalities (\ref{eq:exp1}) and (\ref{eq:exp2}) together with the bound obtained in inequality (\ref{eq:err-bound}) yield,
\begin{equation*}
\Pr[E_p] < k e^{5/2} \max \left \{ (p-k)^{-C'},  \left[ \frac{e(p-k)}{k} \right]^{-kC'} \right \},
\end{equation*}
for $C'=\frac{C-5}{2}$.
\begin{lem} \label{lem-thm2}
For Gaussian perturbation matrices, with $X_{ij} \sim \mathcal{N}(0,1)$ the average error probability that the optimum decoder declares $\F$ is bounded by,
\begin{equation*}
\Pr[\hat{\T}(y,X)=\F|\beta,\T] \leq e^{-\frac{n-k}{2} \log(1+2c\|\btf \|^2) + \frac{d}{2}},
\end{equation*}
with $d=|\T-\F|$ and $c=\frac{3-2\sqrt{2}}{2}$.
\end{lem}
\textbf{Proof.}
The columns of $X_\F$ and $X_{\T-\F}$ are, by definition, disjoint and therefore independent Gaussian random matrices with column spaces spanning random independent $|\F|$- and $|\T-\F|$-dimensional subspaces, respectively.  The Gaussian random vector $\Xbtf$ has i.i.d.\ Gaussian entries with variance $\|\btf\|^2$. 
Therefore, we conclude that, since the random Gaussian vector $\Xbtf$ is projected onto the subspace orthogonal to the random column space of 
$X_\F$, the quantity $\|(I-\Pi_\F)\Xbtf \|^2  /\| \btf\|^2$ 
is a chi-square random variable with $n-k$ degrees of freedom.  
Thus,
\begin{eqnarray*}
\Pr[\hat{\T}(y,X)=\F|\beta,\T] &=& \E_X \left \{ \Pr[\hat{\T}(y,X)=\F|X,\beta,\T]    \right \} \\
&\stackrel{1}{\leq}& \E_X \left \{ e^{-c\|(I-\Pi_\F)\Xbtf \|^2 + \frac{d}{2}}   \right \} \\
&=& \E_{W \sim \chi^2_{n-k}} e^{-c W \|\btf \|^2 + \frac{d}{2}} \\
&\stackrel{2}{=}& e^{-\frac{n-k}{2} \log(1+2c\|\btf \|^2) + \frac{d}{2}}.
\end{eqnarray*}
The first inequality follows from Theorem 1 and the second equality 
comes from the well-known formula for the moment-generating function
of a chi-square random variable,
$\E_{W\sim \chi^2_{n-k}} e^{tW}=(1-2t)^{-\frac{n-k}{2}}$, for $2t<1$.

\section{Conclusion}
In this paper, we examined the probability that the 
optimal decoder declares an incorrect sparsity pattern.
We obtained a sharp upper bound for any generic perturbation matrix,
and this allowed us to calculate the error probability 
in the case of random perturbation matrices. 
In the special case when the entries of the perturbation 
matrix are i.i.d.\ normal random variables, we 
computed an accurate upper bound on the expected 
error probability. Sufficient conditions on exact 
sparsity pattern recovery were obtained, and they
were shown to be stronger than those in previous results 
\cite{Wain07,FRG08,AT08,Karbasi09}.
Moreover, these results match
the corresponding necessary condition presented in \cite{WWR08}. 
An interesting open problem is to extend the sufficient conditions 
derived in this work to non-Gaussian and sparse perturbation matrices.

\section{Acknowledgement} 
The author is grateful to Ioannis Kontoyiannis, Liam Paninski, Xaq Pitkov and Yuri Mishchenko for careful reading of the manuscript and fruitful  discussions.

 \bibliographystyle{ieeetr}
\bibliography{./mybib}

\end{document}